\documentclass[acmtog,nonacm]{acmart}
\usepackage{verbatim}
\usepackage{xcolor}
\usepackage{tabularx}
\usepackage{float}
\usepackage{graphicx}
\usepackage{array}
\usepackage{multirow}
\usepackage{lineno}

\newcommand{\edit}[1]{#1}

\newcolumntype{M}[1]{>{\arraybackslash}m{#1}}

\AtBeginDocument{%
  \providecommand\BibTeX{{%
    \normalfont B\kern-0.5em{\scshape i\kern-0.25em b}\kern-0.8em\TeX}}}

\setcopyright{acmcopyright}
\copyrightyear{2018}
\acmYear{2018}
\acmDOI{10.1145/1122445.1122456}

\acmJournal{TOG}
\acmVolume{37}
\acmNumber{4}
\acmArticle{}
\acmMonth{8}

\begin{document}

%\title{Explainable Artificial Intelligence for Pharmacovigilance \todo{(working title)}}
%\title{Explainable Artificial Intelligence: what does it mean for Pharmacovigilance?}
%\title{Using Explainable Artificial Intelligence to Important fea for Pharmacovigilance}
%\title{Using Explainable Artificial Intelligence to Detect Adverse Drug Outcomes for Pharmacovigilance}
\title{Explainable Artificial Intelligence for Pharmacovigilance: What Features Are Important When Predicting Adverse Outcomes?}
%\title{Training Trustworthy Pharmacovigilance Models with Explainable Artificial Intelligence}
%\title{Using LIME, SHAP, and other Explainable Artificial Intelligence methods for Pharmacovigilance}

\author{Isaac Ronald Ward}
\email{isaacronaldward@gmail.com}
\orcid{}

\affiliation{%
  \institution{School of Population \& Global Health, and Department of Computer Science \& Software Engineering, University of Western Australia}
  \streetaddress{}
  \city{Perth}
  \state{Western Australia}
  \country{Australia}
  \postcode{6009}
}

\author{Ling Wang}
\affiliation{%
  \institution{School of Population \& Global Health, University of Western Australia}
  \streetaddress{}
  \city{Perth}
  \state{Western Australia}
  \country{Australia}
  \postcode{6009}
}

\author{Juan Lu}
\affiliation{%
  \institution{School of Population \& Global Health, University of Western Australia}
  \streetaddress{}
  \city{Perth}
  \state{Western Australia}
  \country{Australia}
  \postcode{6009}
}

\author{Mohammed Bennamoun}
\affiliation{%
  \institution{Department of Computer Science \& Software Engineering, University of Western Australia}
  \streetaddress{}
  \city{Perth}
  \state{Western Australia}
  \country{Australia}
  \postcode{}
}

\author{Girish Dwivedi}
\affiliation{%
  \institution{Cardiology Department Fiona Stanley Hospital, Harry Perkins Institute of Medical Research, Medical School University of Western Australia}
  \streetaddress{}
  \city{Perth}
  \state{Western Australia}
  \country{Australia}
  \postcode{6009}
}

\author{Frank M Sanfilippo$^{*}$}
\affiliation{%
  \institution{School of Population \& Global Health, University of Western Australia}
  \streetaddress{}
  \city{Perth}
  \state{Western Australia}
  \country{Australia}
  \postcode{6009}
}
\thanks{$^{*}$The corresponding author.}

% Removed author addresses
\makeatletter
\let\@authorsaddresses\@empty
\makeatother

%\renewcommand{\shortauthors}{Ward et al.}

% THIS CANNOT EXCEED 300 WORDS:

\begin{abstract}

\subsection*{Background and Objective}

Explainable Artificial Intelligence (XAI) \edit{has been identified as a viable method} for determining the importance of features when making predictions using Machine Learning (ML) models. In this study, we created models that take an individual's health information (e.g. their drug history and comorbidities) as inputs, and predict the probability that the individual will have an Acute Coronary Syndrome (ACS) adverse outcome. 

\subsection*{Methods}

Using XAI, we quantified the contribution that specific drugs had on these ACS predictions, thus creating an XAI-based technique for pharmacovigilance monitoring, using ACS as an example of the adverse outcome to detect. Individuals aged over $65$ who were supplied {M}usculo-skeletal system (anatomical therapeutic chemical (ATC) class M) or {C}ardiovascular system (ATC class C) drugs between $1993$ and $2009$ were identified, and their drug histories, comorbidities, and other key features were extracted from linked Western Australian datasets. Multiple ML models were trained to predict if these individuals would have an ACS related adverse outcome (i.e., death or hospitalisation with a discharge diagnosis of ACS), and a variety of ML and XAI techniques were used to calculate which features --- specifically which drugs --- led to these predictions. 

\subsection*{Results}

\edit{The drug dispensing features for rofecoxib and celecoxib were found to \edit{have a greater than zero contribution} to ACS related adverse outcome predictions (on average)}, and it was found that ACS related adverse outcomes can be predicted with $72\%$ accuracy. Furthermore, the XAI libraries LIME and SHAP were found to successfully identify both important and unimportant features, with SHAP slightly outperforming LIME.

\subsection*{Conclusions}

ML models trained on linked administrative health datasets in tandem with XAI algorithms can successfully quantify feature importance, and with \edit{further development, could potentially be used as pharmacovigilance monitoring techniques.}

\end{abstract}

\begin{CCSXML}
\end{CCSXML}

%\ccsdesc[500]{Computer systems organization~Embedded systems}

\keywords{}

\maketitle
%\onecolumn
%\linenumbers

\section{Introduction}

Artificial Intelligence (AI) and Machine Learning (ML) have recently demonstrated their utility in digital health applications regarding the prediction of outcome events \cite{schmider2019inno, han2019spine}. These techniques use models that learn patterns based on large quantities of data. These models typically demonstrate high predictive power, but they have been criticised as being `black box' algorithms: their internal operations incomprehensible to human operators \cite{das2020xaiopportunities}. In critical decision making domains --- such as healthcare --- the reason for a decision is often as equally important as the decision itself.

In response, Explainable Artificial Intelligence (XAI) has experienced a surge in interest and development. XAI is a field concerned with increasing the explainability and transparency of AI algorithms, by making their influencing variables, complex internal operations, and learned decision making paths interpretable \cite{das2020xaiopportunities, ribeiro2018anchors}. Although popular XAI methods such as `\textbf{L}ocal \textbf{I}nterpretable \textbf{M}odel-Agnostic \textbf{E}xplanations' (\textbf{LIME}) \cite{ribeiro2016lime} and `\textbf{SH}apely \textbf{A}dditive ex\textbf{P}lanations' (\textbf{SHAP}) \cite{lundberg2017shap} have proven to be useful in interpreting black box models \cite{lai2019manyfaces, man2020finance}, more needs to be understood about these methods before they can be adopted in critical decision making domains \cite{dieber2020model}.

In this work, we used pharmacovigilance monitoring as the critical decision making domain which demands transparency and trust. Pharmacovigilance systems aim to recognise pharmaceutical safety issues, and thus impact human well-being, public health, pharmaceutical companies \cite{schmider2019inno}, policy, and regulations at the highest scale. In one case in 2004, the non-steroidal anti-inflammatory drug (NSAID) rofecoxib was withdrawn from global markets based on evidence that it could double the risk of myocardial infarction and stroke if taken for $18$ months or more \cite{debashis2004withdrawal}. At this time, celecoxib, another COX-2 selective inhibitor had its sales reduced by $50\%$ \cite{noauthor2008painful}. Methods which could have quantified and exposed these risks earlier would have significantly improved public health and safety, and both statistical and AI-based methods have been suggested as candidate solutions \cite{lai2017ssa, tsiropoulos2009antiepileptic, hallas1996depresscardio, curtis2008bayesian, wahab2013adrcoxib, wahab2013ssavalid, wahab2016ssahf}. 

The objective of this study was thus two-fold. {{Firstly}}, we used ML to predict the probability of an individual having an Acute Coronary Syndrome (ACS) based adverse outcome. We did this using only administrative health data (Section~\ref{s:ml}). {{Secondly}}, we analysed these ML models with XAI, and confirmed that expected patterns were identified correctly (Section~\ref{s:featimp}). Among other patterns, we expected to find that taking rofecoxib and celecoxib would lead to a disproportionate increase in the predicted probability of an ACS related adverse outcome.

\section{Methods}

\subsection{Datasets}

\edit{The study dataset was prepared by linking data from local core linked administrative datasets (Western Australian Department of Health) and pharmacy dispensing data from the Australian government. All data are de-identified so as to protect sensitive information}. The exposure information is provided by the Pharmaceutical Benefits Scheme (PBS) data, which contains drug dispensing data from PBS-registered pharmacies (community and hospitals). Among other variables, rows in this dataset describe drug dispensing events by their date of supply, quantity, and ATC classification system codes. The outcome events from core datasets of the Western Australian Data Linkage System (WADLS): the Hospital Morbidity Data Collection (HMDC), Emergency Department Data Collection (EDDC), and Deaths Register \cite{holman1999linkeddata}. These datasets contain records of all admissions to public and private hospitals in Western Australia (HMDC), emergency department presentations (EDDC) and deaths, and allow us to identify events by their codes from the International Classification of Diseases 9$^{\textnormal{th}}$ revision clinical modification (ICD-9-CM) and 10$^{\textnormal{th}}$ revision Australian Modification (ICD-10-AM).

\subsection{Data preparation}

To produce data features which can be used for ML, the PBS, HMDC, EDDC, and Deaths Register data were first cleaned, with null entries being dropped entirely. Sex and age are taken from the PBS data, and entries are linked across datasets by the patient's anonymous encrypted identification code. ATC M (musculo-skeletal system) and C (cardiovascular system) class drugs were investigated for this study, as they capture the drugs of interest (rofecoxib and celecoxib) as drugs used by patioents with cardiovascular diseases. We disregard uncommon drugs with less than $10,000$ total dispensing events. 

ACS related hospitalisations and deaths were identified for each individual from the HMDC, EDDC, and Death Register. The study cohort included individuals who were classified as concessional beneficiaries, as their dispensing records within the PBS are more complete than individuals in the general category \cite{paige2015usingpbs}. We additionally limit the study cohort to individuals aged over $65$, as not all concessional beneficiaries are over $65$. Finally, we used data from January $1^{\mathrm{st}}$ $1993$ to December $31^{\mathrm{st}}$ $2009$, which includes the exposure period of interest (2003-2004) with 10-year lookback for comorbidities and followup period to identify outcomes. Critically, it provides us with a large dataset suitable for ML. 

\begin{figure}
    \centering
    \includegraphics[width=0.5\textwidth]{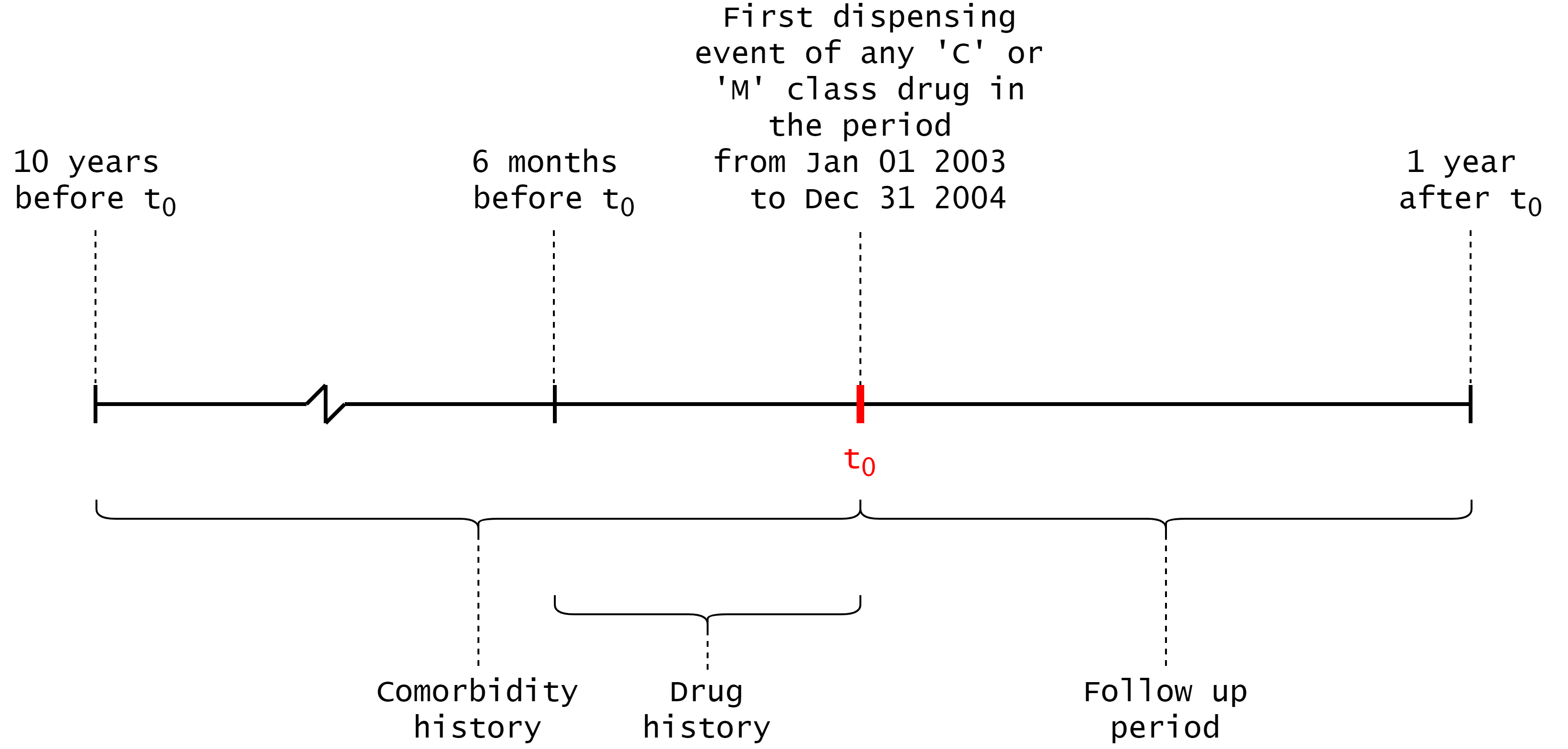}
    \caption{The process by which dispensings, hospitalisation, and death events are converted into drug history and comorbidity history events.}
    \label{fig:timeline}
\end{figure}

The timeline in Figure~\ref{fig:timeline} was used to generate the feature vectors --- the first supply date of a drug is identified via the PBS data, and features are selected around this date. An individual's comorbidity history is defined by the $10$ years prior to the initial supply date, and their drug history is defined by the supply of drugs in the $6$ months prior to the initial supply date. The drug exposure window is defined by the $30$ days following the initial supply date, and the follow up period is the remaining $335$ days in the year after the end of the drug exposure window. 

Hospital admissions or emergency department presentations were considered comorbidities if they were within the $10$ year comorbidity history period, and ACS related hospital admissions or deaths were considered adverse outcomes if they were within the follow up period. These conditions were identified by ICD-9-CM and ICD-10-AM codes. Similarly, any PBS dispensing events for C or M class drugs were added to an individual's drug history if they were within the drug history period. A full list of the conditions considered as adverse outcomes, and the ATC codes of considered drugs are shown in Appendix~\ref{a:lookuptables}.

The $158$-dimension, per patient, per supply feature vectors were constructed by concatenating an individual's sex, age, drug history, comorbidities, and two additional random features. Drug history is represented by $59$ counts of specific C class drug dispensings, and $19$ counts of specific M class drug dispensings, where each element corresponds to a valid ATC drug code which was recorded in the PBS data. The value of each element corresponds to the number of dispensings events the individual had registered for that particular drug in the drug history period. Similarly, hospitalisation comorbidities were represented by $76$ values, where each value corresponds to the number of hospitalisation events that the individual had for that specific condition in the comorbidity period. The random features were added because by definition they have no predictive power. This aids in quantifying the models' tendency to overfit, and in quantifying the feature importance methods' ability to detect truly important features. A random integer (in the range $0$--$158$) and floating point number (in the range $0$--$1$) are appended to each feature vector. \edit{Note that every feature in the feature vector is represented numerically, and that scikit-learn interprets all features as continuous numeric variables with threshold-based decision boundaries --- thus preventing the possibility of bias due to data type.}

The label for a feature vector was a $0$ or $1$ (one-hot encoding), denoting if an individual suffered an ACS related adverse outcome or death during the follow up period ($1$) or not ($0$). \edit{We removed all instances which had identical feature vectors but different labels, as these cases lacked the input information required to discriminate between the two possible outcomes.} This limitation is inherent in administrative datasets; such datasets do not capture all the information about an individual that describes their full clinical history and presentation. 

The resulting linked dataset is imbalanced: the number of instances which did not have ACS related adverse outcomes (the negative class) outnumbered those who did (the positive class) by a ratio $6.94$:$1$. For the training set, we used 70\% of the instances and randomly undersampled from the negative class population until the training dataset was exactly equal, totalling $278,608$ instances \cite{lemaitre2017imbalanced}. \edit{Random undersampling approaches have been proven to be suitable for dealing with minority classes with relative class sizes that were even smaller than ours (our relative minority class size is $14.41$\%, and $0.1$\% minority classes have been trained with undersampling with minimal performance losses in \cite{hasanin2018effectsrus}).} Having a balanced training set will prevent the ML models from learning to predict the larger class in lieu of identifying meaningful patterns. \edit{We used $4$-fold cross validation when training and testing, with a ratio of 30\% of the total instances reserved for the testing set. Performance measures were averaged across folds.}

\subsection{Machine Learning}
\label{s:ml}

We used Decision Tree (DT) based classification models: \textbf{R}andom \textbf{F}orests (RFs), \textbf{E}xtra random \textbf{T}rees (ETs), and e\textbf{X}treme \textbf{G}radient \textbf{B}oosting machines (XGB) \cite{breiman2001rf,geurts2006extratrees,chen2016xgb}. These tree based models benefit from being both widely accepted and used, as well as having accelerated feature importance value computations \cite{lundberg2017shap} (see Section~\ref{s:featimp}). Moreover, using a variety of proven ML models allows us to investigate if our results are not model dependent.

\begin{figure}
    \centering
    \includegraphics[width=0.5\textwidth]{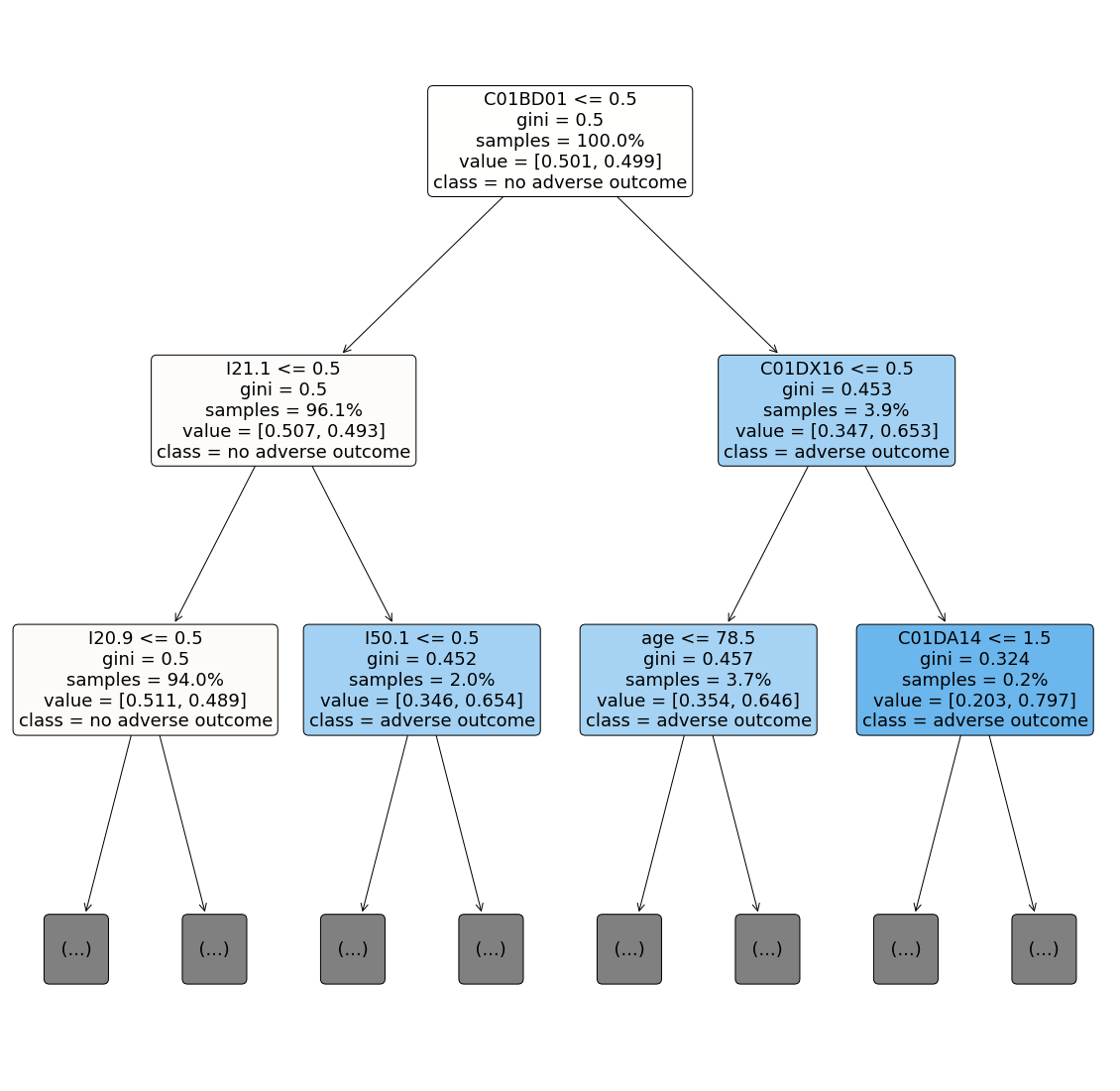}
    \caption{Visualisation of a portion of a single DT from the trained RF used in this study. At each node, a feature and threshold is selected, for example: in the top-most node, the decision is based on whether or not an individual was dispensed C01BD01 (dofetilide) in their drug history. Other drug dispensing features, comorbidities, and the individual's age are also used as decision features in this image. The Gini impurity, the ratio of samples which make it to that node (samples), the ratio of samples which are split into each branch of that node (value), and the majority class (class) rendered. The DT is cut off after the third ply in this figure, as the actual DT's depth is over $30$ in this case. DT, Decision Tree; RF, Random Forest.}
    \label{fig:dt}
\end{figure}

A \textbf{DT} is a representation of an algorithm that follows a tree-like structure (see Figure~\ref{fig:dt}). In ML, DTs can be generated from labelled datasets --- during each training step, a condition is based on some selected feature (either randomly or based on some measure of optimality), and the data is split into subsets based on the outcome of this condition. Each splitting point creates new nodes, and this process continues recursively until some stopping criteria has been met (e.g. accuracy, node limits). One measure of optimality is \textit{Gini impurity}, which measures the probability of an instance being classified incorrectly when classifying it based on the class distribution of the dataset. 

\edit{It is also possible to identify the most important features by inspecting the DTs directly; the features that are used as splitting criteria in shallow nodes discriminate between feature vectors more effectively (i.e., these features have the highest \textit{information gain}). In practice, we find that single DTs are prone to overfitting due to the small samples that occur near the tree's leaf nodes. As such, all of our feature importance analyses are based on ensembles of trees, rather than single trees, which will ensure that the calculated feature importance scores are less variant and more indicative of the global patterns in the data. }

An \textbf{RF} is an ensemble architecture which bags multiple DTs to produce more stable predictions. DTs with high complexity often overfit, whereas RFs create subsets of random features and build a greater number of smaller DTs using these subsets. This has a tendency to increase variance and thus reduce overfitting \cite{breiman2001rf}. Furthermore, \textbf{ET}s further increase the variance by also randomly choosing the decision threshold at each node in each DT \cite{geurts2006extratrees}. \textbf{XGB}s are another bagging approach to DTs, but in this case boosting is used to alter the evaluation criteria for each DT. These errors are minimised via gradient descent. XGBs have numerous performance optimisations which suit them for the purpose of this study \cite{chen2016xgb, shwartz2021tabular}.

To optimally train these models we performed $128$ rounds of Bayesian hyperparameter optimisation over a distribution of valid parameters ({see Appendix~\ref{a:hyper} for the descriptions of these searches}) \cite{snoek2012bayesianopt}. In each round, a configuration of parameters is selected and the resulting model is trained and tested using $4$-fold cross validation \cite{roth2015deeporgan4fold}. Performance measures were averaged across folds. The most performant model from the $128$ trials is selected and its feature importance is analysed using both traditional methods and XAI \edit{(the results of which are averaged across the three models and presented in Figure~\ref{tbl:table_of_feat_imps})}.

\subsection{Feature importance}
\label{s:featimp}

We used ML-based feature importance methods to understand the contribution of certain features to the model's prediction. From a pharmacovigilance perspective, we want to know which drug(s) have the highest association with the outcome. The traditional methods that we employed were measures of \textbf{M}ean \textbf{D}ecrease of \textbf{I}mpurity (MDI), and \textbf{M}ean \textbf{D}ecrease in \textbf{A}ccuracy (MDA), and the XAI-based methods are LIME and SHAP. 

\textbf{MDI} is based on the depth of certain nodes in a tree-based model's DTs. If a node near the top of the DT uses some criterion based on a given feature, then that feature must contribute to the final decision for a larger fraction of samples than a feature which is used lower in the DT. This fraction can be used as an estimate of a feature's relative importance. MDI is calculated by combining the decrease in Gini impurity with this relative importance \cite{louppe2015understanding}. \edit{We used scikit-learn's implementation of MDI in this study}. 

\textbf{MDA} is defined as the decrease in model accuracy on the test set when a given feature is randomised or permuted \cite{breiman2001rf, altmann2010permutation}. The drop in accuracy indicates how much the model depends on the given feature, and thus is an estimate of feature importance. MDA benefits from being model agnostic, unlike MDI measures which need to be performed on tree-based models. \edit{Additionally, MDA does not suffer from a bias towards high cardinality features (as MDI does)}. The method does still suffer from some bias however --- when two or more features are highly correlated, permuting one feature does not restrict the model's access to it, as it can still be accessed via the correlated feature(s). 

\textbf{L}ocal \textbf{I}nterpretable \textbf{M}odel-agnostic \textbf{E}xplanations (\textbf{LIME}) \cite{ribeiro2016lime} generate local surrogate models to explain individual predictions produced from ML architectures. LIME takes a single instance, and locally \textit{perturbs} the feature values to other valid values based on the dataset. These perturbed values are then fed back into the trained model. This creates a new dataset of input to output mappings, which an interpretable model is trained on. During the training of this local surrogate model, instances are weighted by the distance of the perturbed feature to the single feature of interest. The result is a ML model which is explainable and has a high local fidelity: it approximates the black box model locally in feature space, but not globally. LIME has been noted to increase model interpretability on tabular data \cite{dieber2020model}, but relies on the correct definition of the local neighborhood and is thus highly dependent on kernel parameters \cite{laugel2018definelocal}.

\textbf{SH}apley \textbf{A}dditive ex\textbf{P}lanations (\textbf{SHAP}) \cite{lundberg2017shap} assigns each input feature an importance value for a given prediction, based on principles from cooperative game theory. For any given feature vector, SHAP takes a single feature's value and replaces it with a sampled value from the dataset. The model of interest makes a prediction for this augmented feature vector and the difference in output values --- the marginal contribution --- is noted. This sampling process can be repeated to improve our estimates of marginal contribution. This is repeated for all possible coalitions, and the resulting average marginal contributions to each coalition are the Shapley values. SHAP can approximate these values accurately and quickly for tree-based models.

\section{Results}

\subsection{Predicting per-patient adverse outcomes}

Table~\ref{tab:modelscore} presents the results from the hypertuning experiments for each of the model classes. The hypertuning procedure converges each model to similar levels of accuracy and Area Under the Receiver Operating Characteristic (AUC). The XGB classifier outperforms the RF classifier, which outperforms the ET classifier. 

\begin{table}[H]
\centering
\caption{The test results after training the RF, ET and XGB models used in this study. For each model, a Bayesian hyperparameter optimisation search with $128$ iterations was performed. Macro-average precision was used here, as this measure is \textbf{insensitive} to the imbalance of classes in the test dataset. }
\begin{tabularx}{0.45\textwidth}{ l  c  c  c  c }
    \toprule
     &   & Accuracy & Macro-avg.  & Macro-avg.  \\ 
     Model\hspace{5mm} & AUC &  (\%) & Precision (\%) & Recall (\%)  \\ 
    \toprule
    %RFC & 0.69 & 0.69 & 0.69 & 0.69 \\  
    RF  & 0.70 & 71 & 67 & 70 \\  
    %RFC (grid. search) & 0.69 & 0.69 & 0.69 & 0.69 \\  
    \midrule
    %ETC & 0.69 & 0.69 & 0.69 & 0.69 \\  
    ET  & 0.69 & 70 & 66 & 69 \\  
    %ETC (grid. search) & --- & --- & --- & --- \\  
    \midrule
    %XGB & 0.67 & 0.67 & 0.67 & 0.67 \\  
    XGB  & 0.72 & 72 & 68 & 72 \\  
    %XGB (grid. search) & --- & --- & --- & --- \\  
    \bottomrule
\end{tabularx}
\caption*{AUC, Area under the ROC Curve; RF, Random Forest; ET, Extra Trees (classifier); XGB, Extreme Gradient Boosting.}
\label{tab:modelscore}
\end{table}

\subsection{Generating feature importance for pharmacovigilance}

We analysed the models' feature importance scores using the four measures described in Section~\ref{s:featimp}. The per-feature importance scores as determined by MDI, MDA, LIME, and SHAP are plotted in Figure~\ref{tbl:table_of_feat_imps}. \edit{There is an important distinction here: for MDI and MDA measures, we showed the importance of any feature when making either an adverse outcome \textit{or a non adverse outcome prediction}, as these measures do not give per-instance feature contributions. For LIME and SHAP, we present results for specifically making adverse outcome predictions, and we note that it is possible for a feature to attain on-average negative contribution to an ACS related adverse outcome prediction (these features `protect' a patient from having an ACS related prediction).} 

We observed that age and sex are almost always attributed with a high importance --- in MDI and MDA especially, but less so under LIME analysis. There are repeating peaks and patterns of importance across MDI, MDA, LIME, and SHAP in C class drug history, M class drug history, and ICD-10-AM codes (see Figure~\ref{tbl:table_of_feat_imps} and Appendix~\ref{a:lookuptables} for more detail). Rofecoxib and celecoxib are the most important M class drug history features under MDI and MDA analysis, but not so under LIME analysis. Under SHAP analysis, rofecoxib and celecoxib are the second and third M class drug history features that on average contribute to an adverse outcome prediction respectively, being slightly behind M05BA04 (alendronic acid), with the feature importance scores of rofecoxib and alendronic acid being $1.16\times10^{-4}$ and $1.19\times10^{-4}$ respectively. We note the random features' importances under MDI analysis, which is a result of the method's bias towards high cardinality features \cite{altmann2010permutation}.

We further analysed the feature importance rankings at the model level by displaying the key feature importance rankings in Table~\ref{tab:rankings}. The results largely reflect the average results, but the XGB classifier notably disfavors random features under MDI analysis, and favors the random floating point feature under MDA and SHAP analysis. \edit{We note that different model types may correlate to the features that are considered important --- even across feature importance methods. This is potentially due to inductive bias in the learning algorithm's design. Indeed, RFs and ETs are algorithmically similar, with the only key difference being the randomisation of selected thresholds and the sampling methods used in ETs.}

\edit{We also provide the per-model rankings of the M class drug features in Table~\ref{tab:permodelmclass}, and we observe that the XGB model disagrees with the RF and ET models under MDI analysis, but agrees with the RF and ET models under MDA analysis. This is potentially due to the shortcomings of MDI compared to MDA, which are discussed in the next section. LIME analysis suffers from high variance and inconsistency in the results, potentially due to the method’s high local fidelity. SHAP analysis results are largely consistent, apart from the RF not identifying celecoxib dispensing as an important feature whereas the ET and XGB models both do.}

\begin{table}[h]
\centering

\caption{Rankings of features as measured by MDI, MDA, LIME, and SHAP. Key confounders sex and age are expected to be ranked highly, and random features to be ranked lower.}
\begin{tabularx}{0.45\textwidth}{ l  l  >{\centering\arraybackslash}m{0.89cm} >{\centering\arraybackslash}m{0.89cm} >{\centering\arraybackslash}m{0.89cm} >{\centering\arraybackslash}m{0.89cm} }
    \toprule
    \multirow{2}{*}{Measure}              & \multirow{2}{*}{Feature}        & \multicolumn{4}{c}{Overall ranking (out of $158$)}  \\ \cline{3-6}
           &  & RF & ET & XGB & Avg. \\ 
    \toprule
    
    % For MDI
    \multirow{4}{*}{MDI} 
    & sex & 8 & 4 & 5 & 5  \\ 
    & age & 1 & 1 & 20 & 1  \\ 
    & random float & 3 & 20 & 113 & 8  \\   
    & random int & 4 & 21 & 116 & 10  \\  
    \midrule
    
    % For MDA
    \multirow{4}{*}{MDA} 
    & sex & 4 & 2 & 5 & 2  \\ 
    & age & 1 & 1 & 1 & 1  \\ 
    & random float & 157 & 158 & 43 & 157  \\   
    & random int & 158 & 157 & 158 & 158  \\  
    \midrule
    
    % For LIME
    \multirow{4}{*}{LIME} 
    & sex & 29 & 37 & 46 & 40  \\ 
    & age & 43 & 53 & 63 & 57  \\ 
    & random float & 42 & 56 & 72 & 61  \\   
    & random int & 75 & 101 & 84 & 86  \\  
    \midrule
    
    % For SHAP
    \multirow{4}{*}{SHAP} 
    & sex & 7 & 2 & 7 & 5  \\ 
    & age & 1 & 1 & 13 & 10  \\ 
    & random float & 155 & 128 & 20 & 35  \\   
    & random int & 150 & 114 & 126 & 139 \\  
    \bottomrule
    
\end{tabularx}
\caption*{The average score is not the average rank, but the rank of the average importance for all models across that measure, as rendered in Figure~\ref{tbl:table_of_feat_imps}. The feature with the highest importance is ranked as $1$, and the lowest as $158$. MDA, Mean Decrease in Impurity; MDA, Mean Decrease in Accuracy; LIME, Local Model Agnostic Explanations; SHAP, Shapley Additive Explanations; RF, Random Forest; ET, Extra Trees (classifier); XGB, Extreme Gradient Boosting}
\label{tab:rankings}
\end{table}

\begin{table}
\centering
\caption{The rankings of the M class drug history features specifically, as measured by MDI, MDA, LIME, and SHAP. The feature with the highest importance is ranked a $1$, and the lowest as $19$. The drugs of interest to this study --- celecoxib and rofecoxib --- have their importances across all models shown. The average score is not the average rank, but the rank of the average importance for all models across that measure, as rendered in Figure~\ref{tbl:table_of_figures}.}
\begin{tabularx}{0.45\textwidth}{ l  l  >{\centering\arraybackslash}m{0.89cm} >{\centering\arraybackslash}m{0.89cm} >{\centering\arraybackslash}m{0.89cm} >{\centering\arraybackslash}m{0.89cm} }
    \toprule
    \multirow{2}{*}{Measure}              & \multirow{2}{*}{Feature}        & \multicolumn{4}{c}{M class drug ranking (out of $19$)}  \\ \cline{3-6}
           &  & RF & ET & XGB & Avg. \\ 
    \toprule
    
    % For MDI
    \multirow{2}{*}{MDI}  
    & celecoxib & 1 & 2 & 18 & 1  \\ 
    & rofecoxib & 2 & 1 & 17 & 2  \\   
    \midrule
    
    % For MDA
    \multirow{2}{*}{MDA}  
    & celecoxib & 1 & 2 & 1 & 1  \\ 
    & rofecoxib & 2 & 3 & 3 & 2  \\  
    \midrule
    
    % For LIME
    \multirow{2}{*}{LIME} 
    & celecoxib & 15 & 9 & 10 & 12  \\ 
    & rofecoxib & 4 & 11 & 12 & 9  \\  
    \midrule
    
    % For SHAPHi
    \multirow{2}{*}{SHAP} 
    & celecoxib & 10 & 4 & 3 & 3  \\ 
    & rofecoxib & 1 & 2 & 1 & 2  \\   
    \bottomrule
    
\end{tabularx}
\caption*{MDA, Mean Decrease in Impurity; MDA, Mean Decrease in Accuracy; LIME, Local Model Agnostic Explanations; SHAP, Shapley Additive Explanations; RF,Random Forest; ET, Extra Trees (classifier); XGB, Extreme Gradient Boosting}
\label{tab:permodelmclass}
\end{table}

\begin{table*}
    \centering
    \begin{tabular}{M{1.0\textwidth}}
    \includegraphics[width=0.98\textwidth]{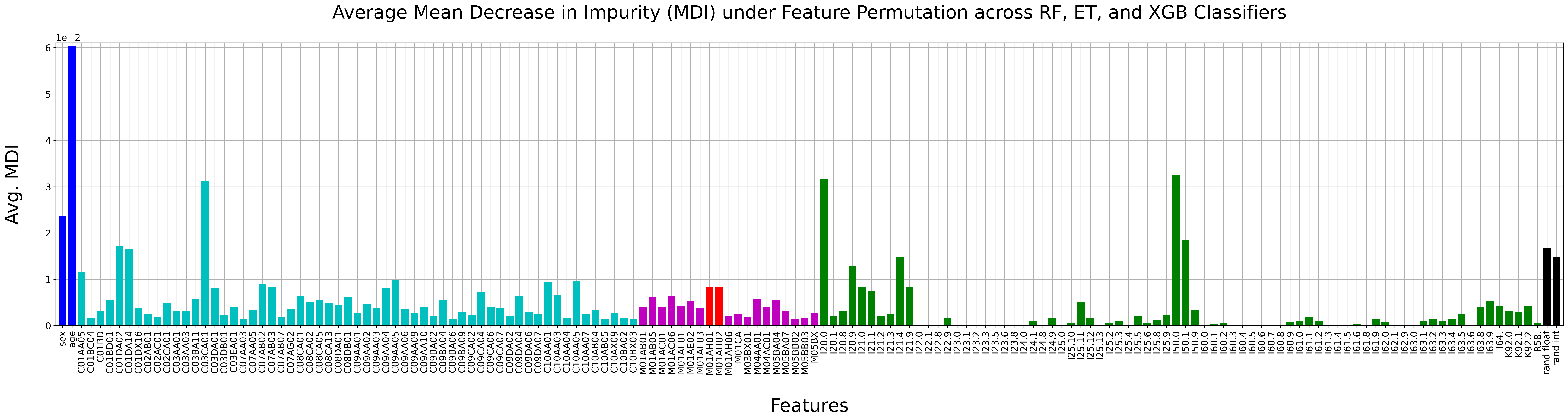} \\
    \includegraphics[width=0.98\textwidth]{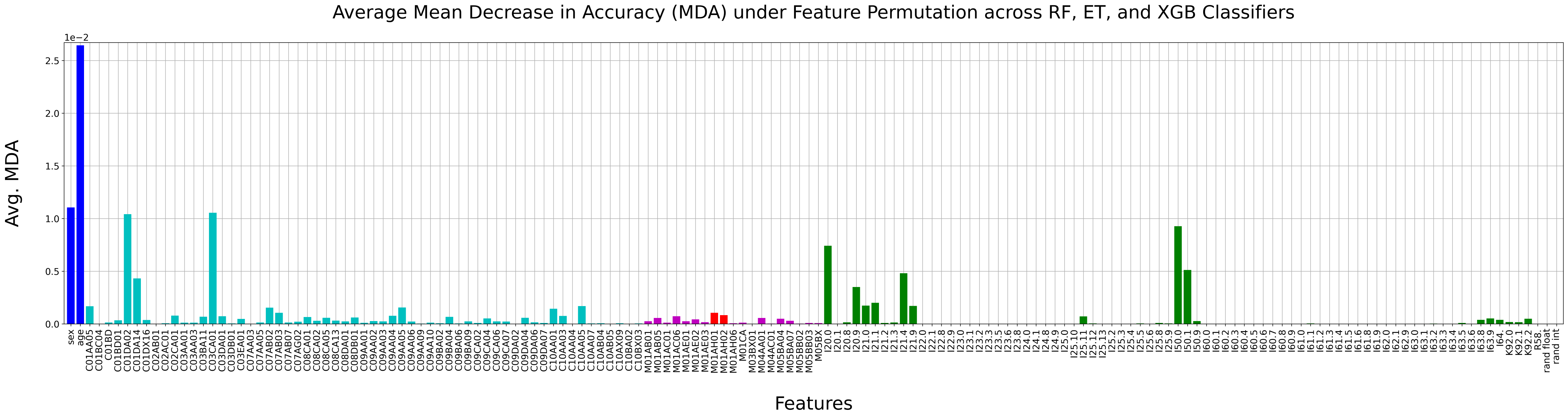} \\
    \includegraphics[width=0.98\textwidth]{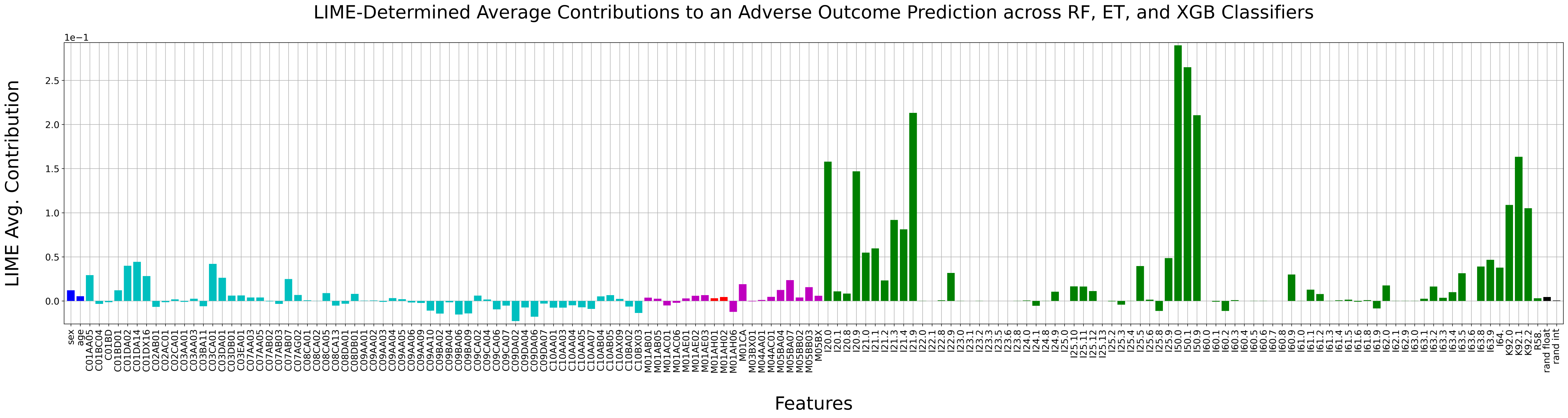} \\
    \includegraphics[width=0.98\textwidth]{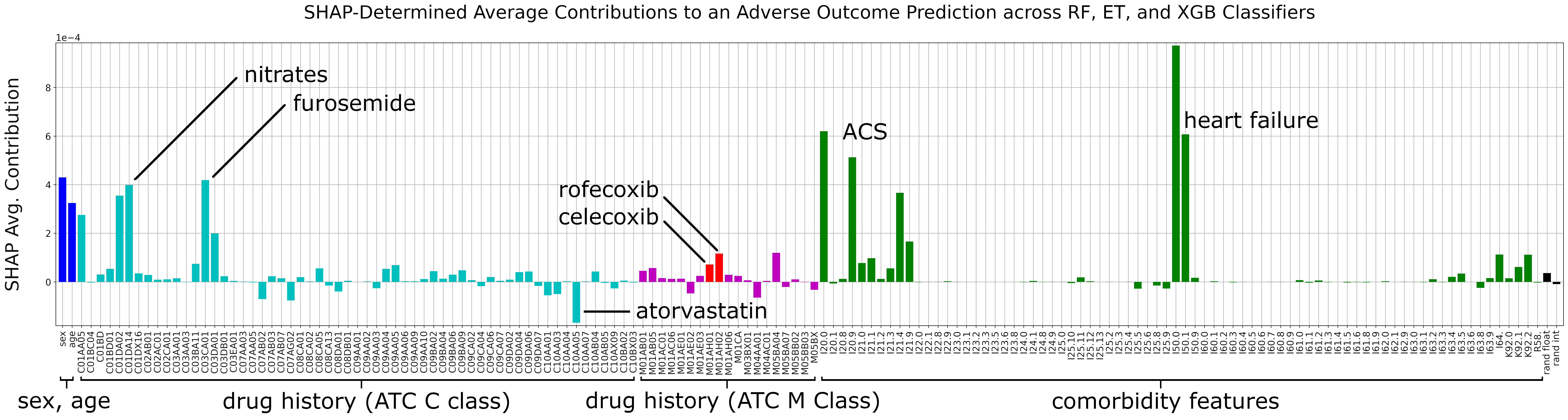}
    \end{tabular}

    \captionof{figure}{Names of the features used to predict an ACS related adverse outcome, with \edit{feature importance scores (MDI, MDA, LIME, and SHAP) averaged across the RF, ET, and XGB models (to reduce cross-model variance)}. Sex and age are rendered in dark blue, C class drug history features in light blue, M class drug history features in magenta (with celecoxib and rofecoxib highlighted in red), comorbidities in green, and random features in black. In all importance measures other than LIME, the key confounders sex and age are identified as important features, and certain C class drugs, M class drugs, and ICD-10-AM codes are consistently identified as important in predicting an ACS related to the adverse outcome. Note the high importance of celecoxib (M01AH01) and rofecoxib (M01AH02) dispensings when compared to other M class drugs under MDI, MDA, and SHAP analysis (this is shown numerically in Table~\ref{tab:permodelmclass}). All features are described in Appendix~\ref{a:lookuptables}. Image best viewed in colour. MDA, Mean Decrease in Impurity; MDA, Mean Decrease in Accuracy; LIME, Local Model Agnostic Explanations; SHAP, Shapley Additive Explanations; RF, Random Forest; ET, Extra Trees (classifier); XGB, Extreme Gradient Boosting; ACS, acute coronary syndrome; ICD, International Classification of Disease. }
    \label{tbl:table_of_feat_imps}
\end{table*}

\begin{comment}
    \begin{figure*}
        \includegraphics[width=0.97\textwidth]{figs/full/mdi.png}
        \label{fig:mdi}
        
        %\caption{This is a tiger.}
    \end{figure*}
    \begin{figure*}
        \includegraphics[width=0.97\textwidth]{figs/full/mda.png}
        \label{fig:mda}
      
        %\caption{This is a tiger.}
    \end{figure*}
    \begin{figure*}
        \includegraphics[width=0.97\textwidth]{figs/full/lime.png}
        \label{fig:lime}
      
        %\caption{This is a tiger.}
    \end{figure*}
    \begin{figure*}
        \includegraphics[width=0.97\textwidth]{figs/full/shap2.png}
        \label{fig:shap}
      
        \caption{This is a tiger.}
    \end{figure*}
\end{comment}

\subsection{Stratified analysis using Explainable Artificial Intelligence}

The XAI methods LIME and SHAP can extract per-instance prediction contributions, which allowed us to focus our analysis on the \textit{values} of certain features (i.e., we can investigate local, rather than global importance scores). We divided the test datasets into four distinct strata, depending on whether or not an individual was or was not dispensed celecoxib or rofecoxib. For this analysis, we focussed on the M class drugs and present these findings in Figure~\ref{tbl:table_of_figures}. We observed that rofecoxib and celecoxib dispensings increases the likelihood of an adverse outcome prediction, and that \textit{not} having any dispensings of either drug decreased the likelihood. In other words: having rofecoxib and celecoxib dispensings on average increased the likelihood of an adverse outcome prediction, and having no such dispensings of these drugs on average decreased or had little effect on the likelihood (in both LIME and SHAP).

\begin{table*}
    \caption*{The Effect of Being Dispensed M01AH01 (celecoxib) and M01AH02 (rofecoxib) on the Likelihood of a Predicted ACS Related Adverse Outcome}
    \vspace{-5mm}
    \centering
    \begin{tabular}{p{0mm} p{0.41\textwidth} p{5mm} p{0mm} p{0.41\textwidth}}
        
        \vspace{0mm}a) & \raisebox{-\totalheight}{\includegraphics[width=0.41\textwidth]{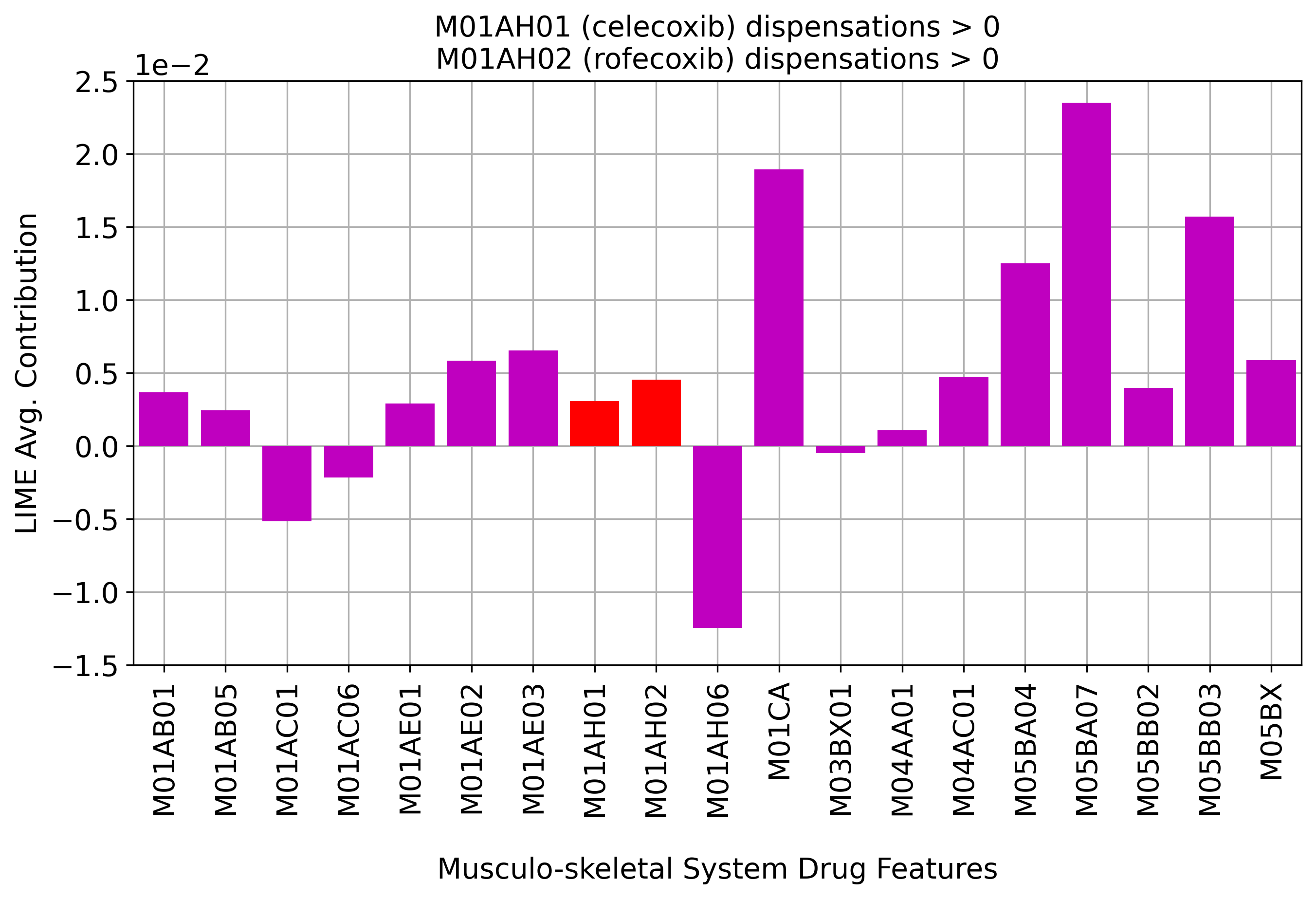}} & & \vspace{5mm} & \raisebox{-\totalheight}{\includegraphics[width=0.41\textwidth]{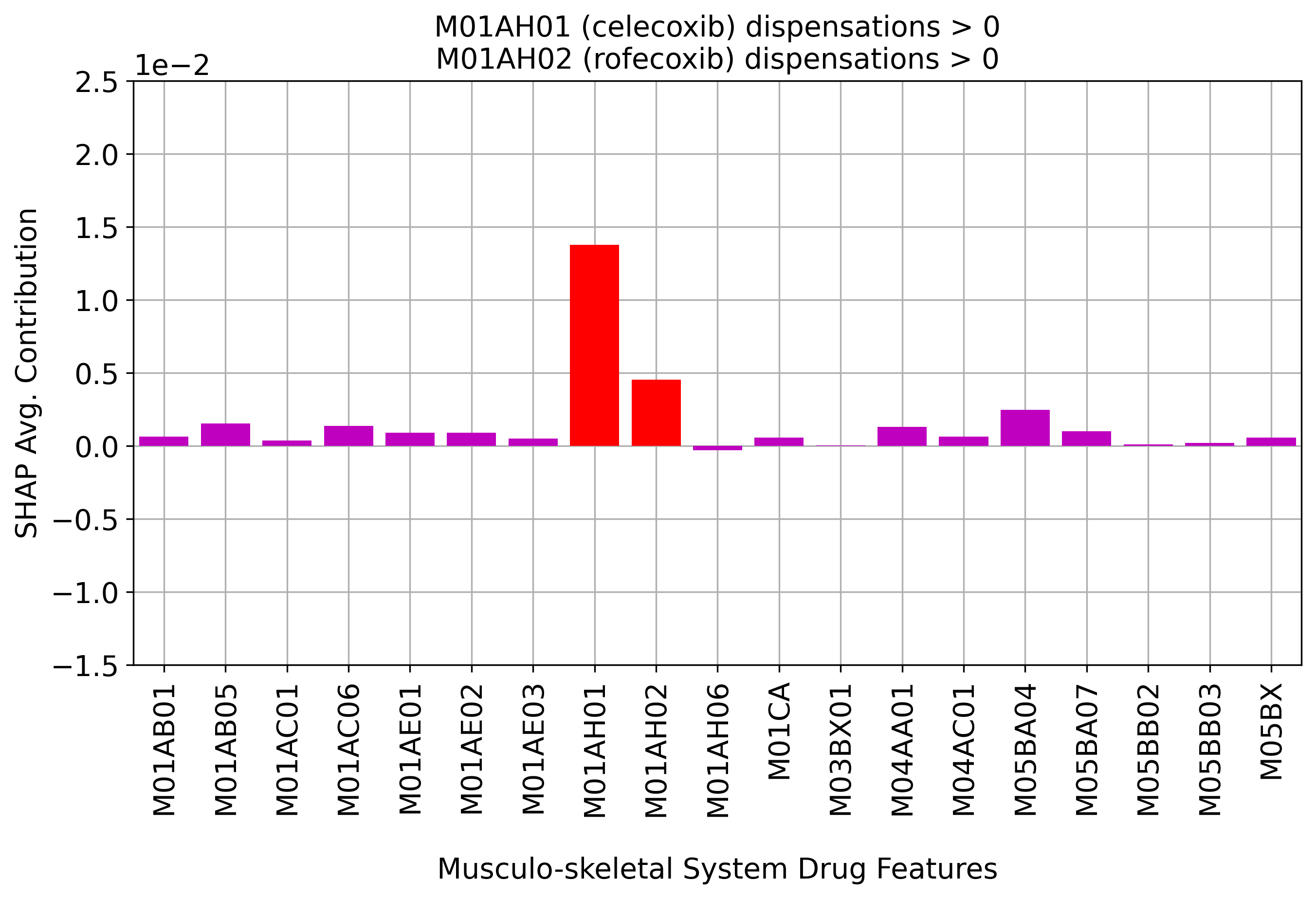}} \\
        
        \vspace{0mm}b) & \raisebox{-\totalheight}{\includegraphics[width=0.41\textwidth]{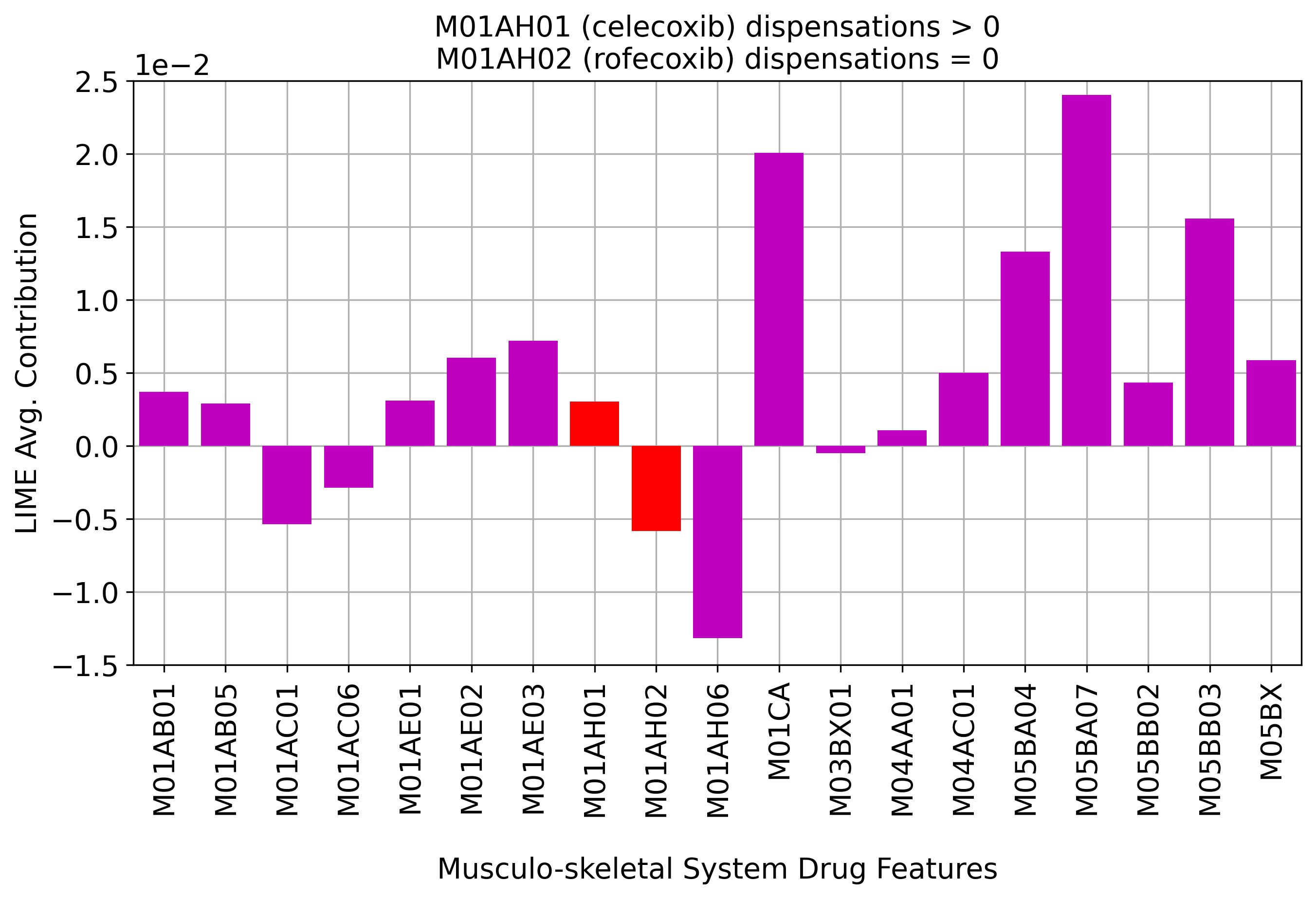}} & & \vspace{5mm} & \raisebox{-\totalheight}{\includegraphics[width=0.41\textwidth]{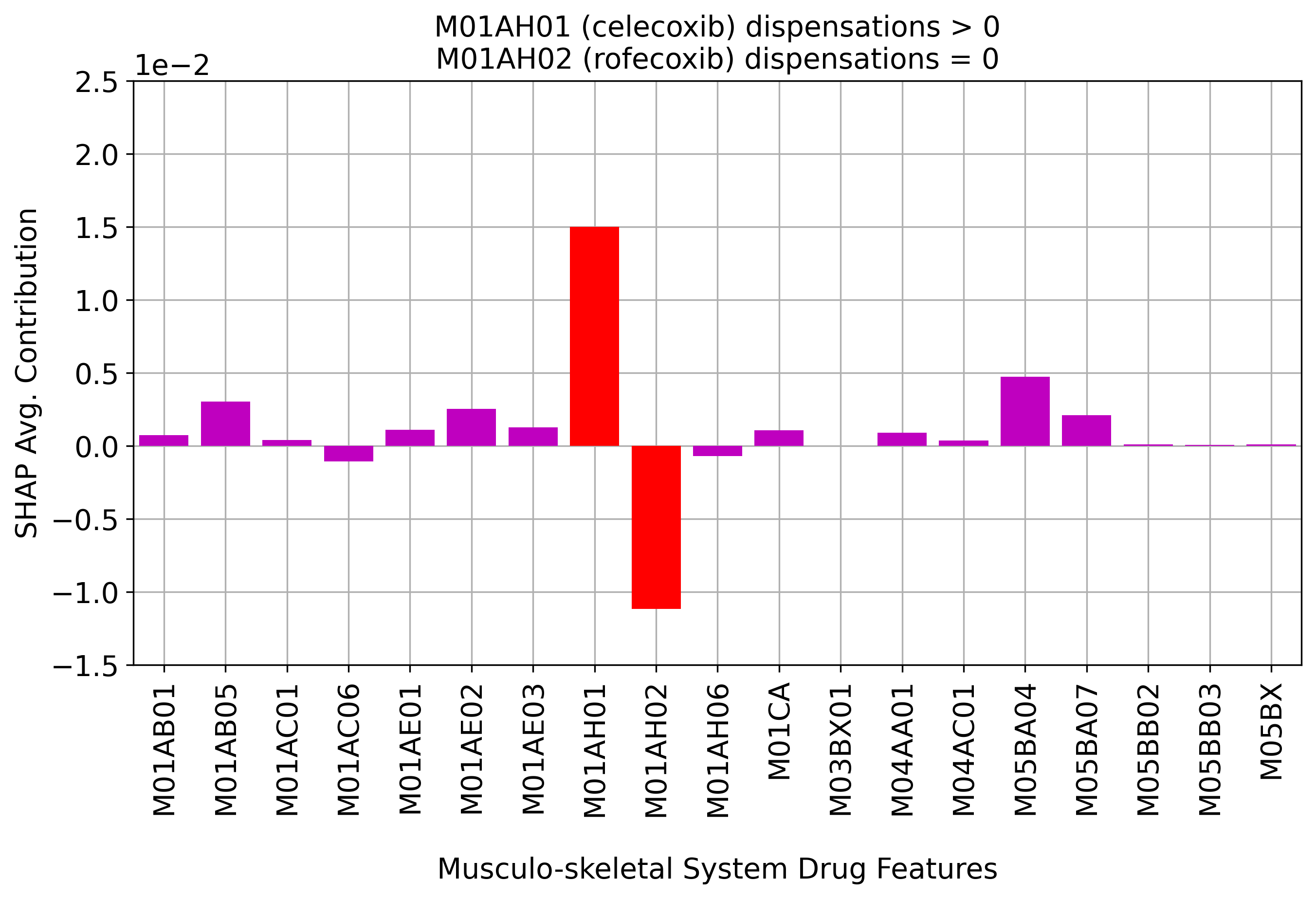}} \\
        
        \vspace{0mm}c) & \raisebox{-\totalheight}{\includegraphics[width=0.41\textwidth]{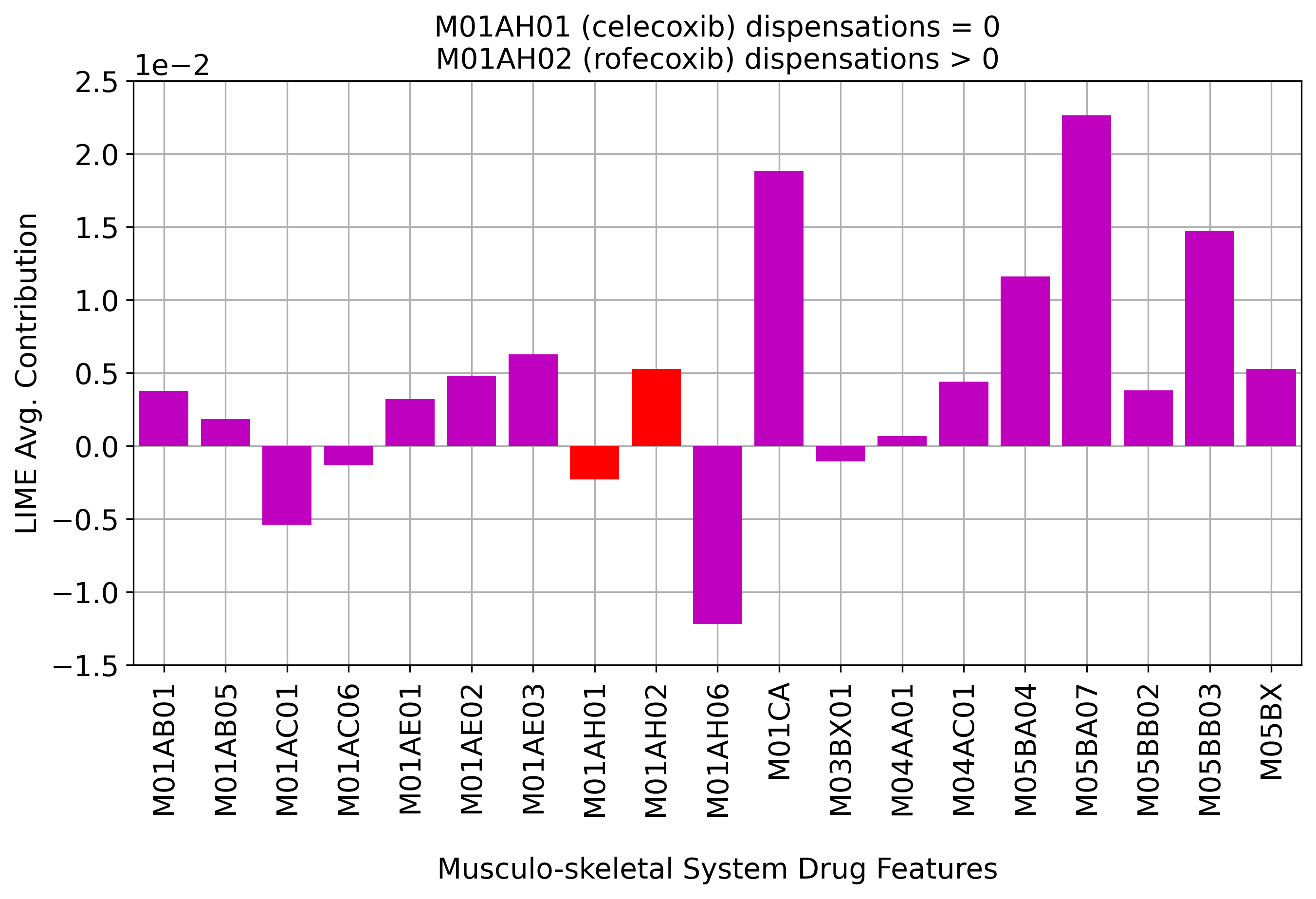}} & & \vspace{5mm} & \raisebox{-\totalheight}{\includegraphics[width=0.41\textwidth]{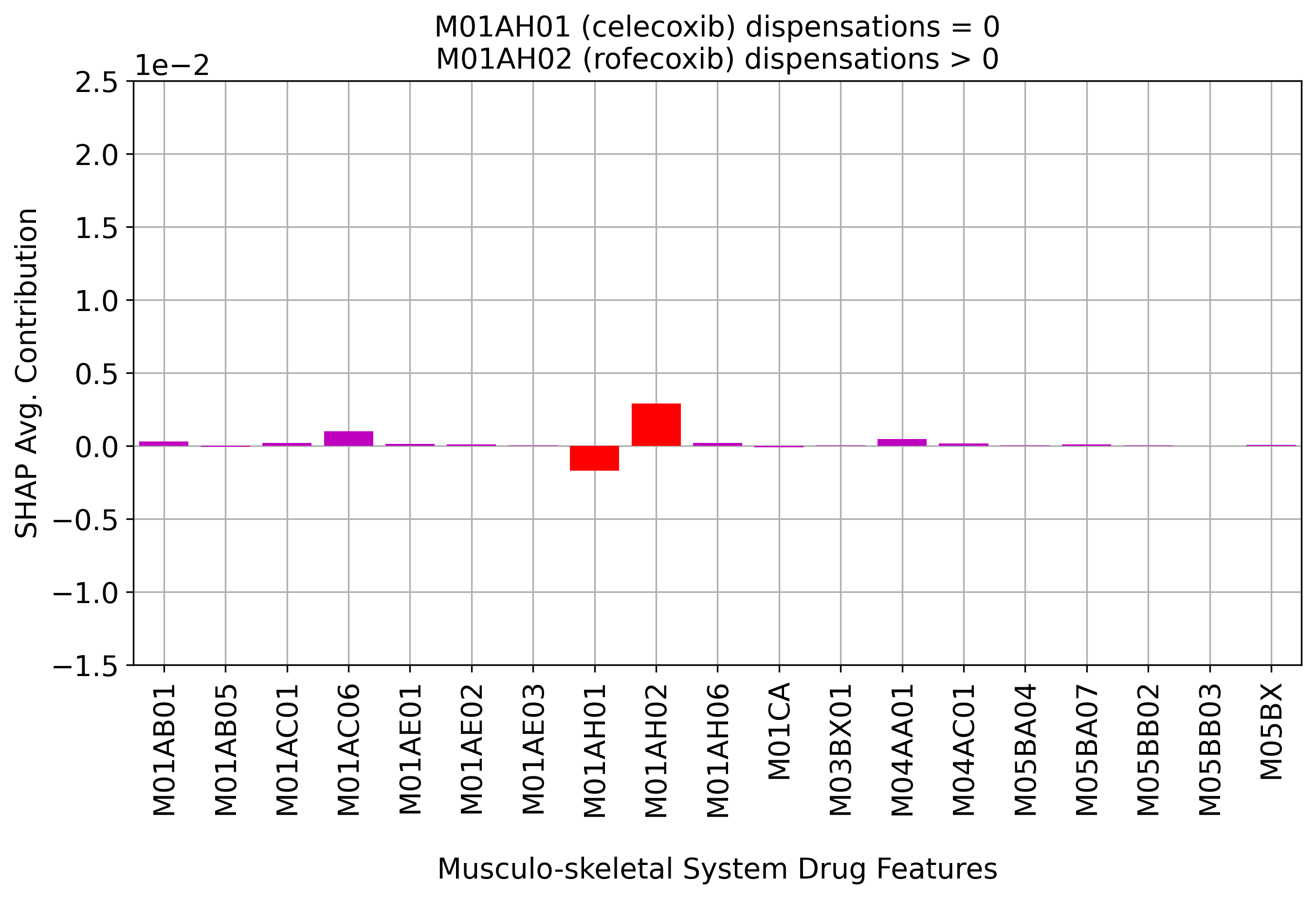}} \\
        
        \vspace{0mm}d) & \raisebox{-\totalheight}{\includegraphics[width=0.41\textwidth]{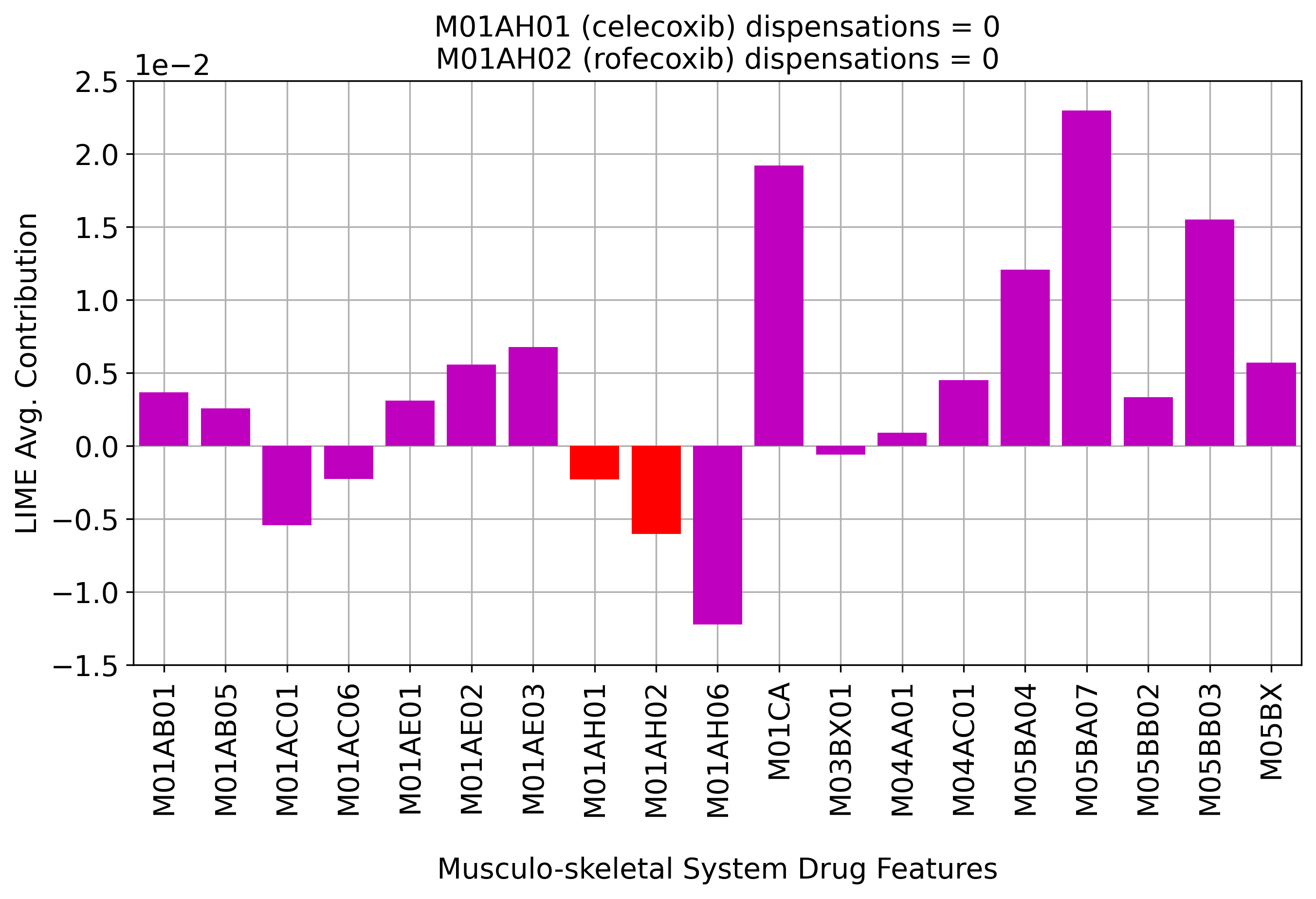}} & & \vspace{5mm} & \raisebox{-\totalheight}{\includegraphics[width=0.41\textwidth]{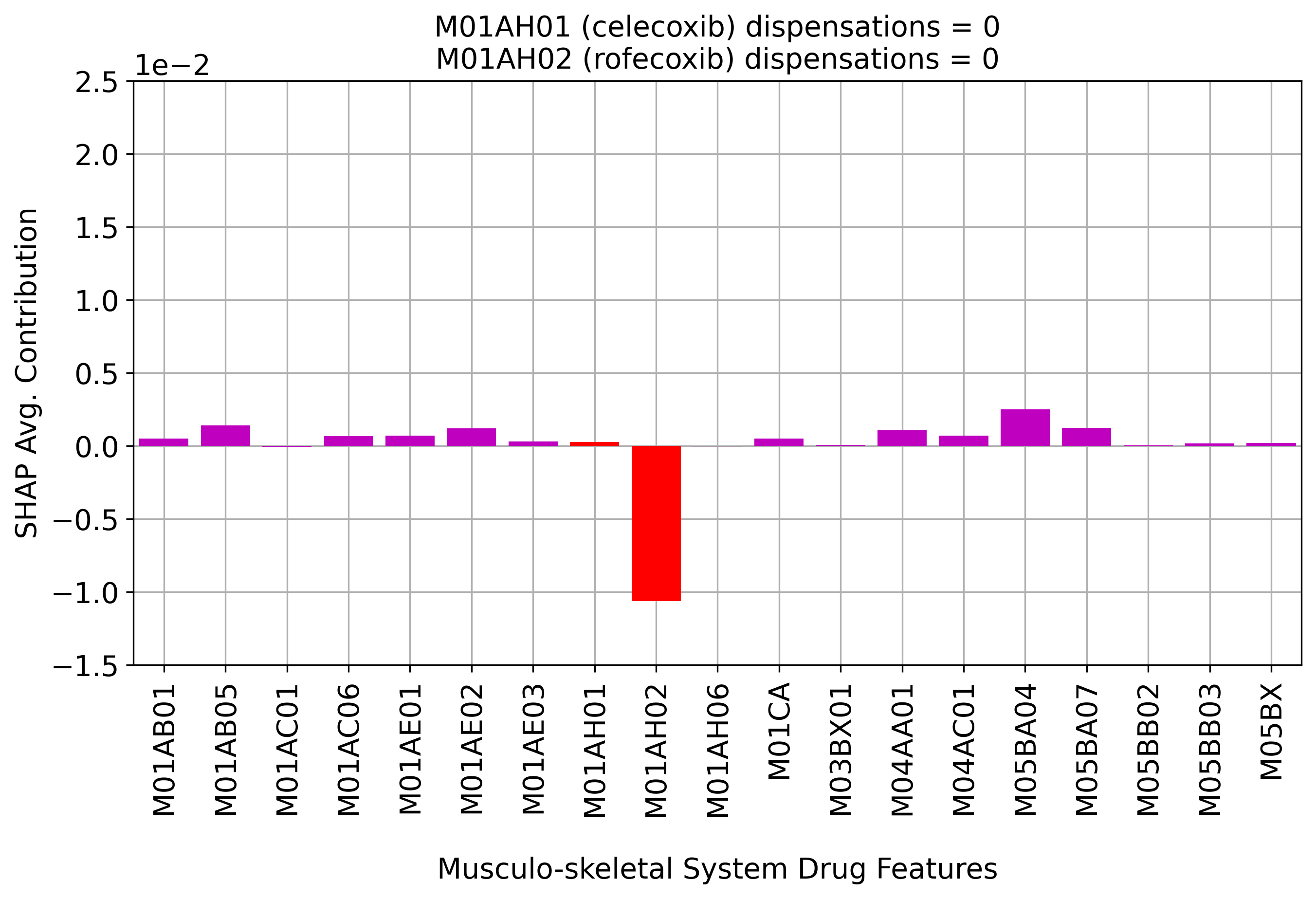}} \\
        
    \end{tabular}

    \captionof{figure}{Dividing the test datasets into strata where an individual was a) dispensed celecoxib and rofecoxib, b) dispensed celecoxib and not rofecoxib, c) not dispensed celecoxib and dispensed rofecoxib, and d) where neither drug was dispensed. The contribution of M class drug history features under LIME analysis (left column) and SHAP analysis (right column) are presented. If individuals have been dispensed celecoxib or rofecoxib this on average resulted in an increased likelihood of a predicted ACS related adverse outcome. If an individual had not been dispensed rofecoxib or celecoxib, then this on average resulted in a near zero or \textit{decreased} likelihood of a predicted ACS related adverse outcome. Note that, on average, the \textit{other} M class drug dispensing events did not vary across these subsets, so we do not expect their feature scores to vary across rows a), b), c), and d). Celecoxib (M01AH01) and rofecoxib (M01AH02) are shown in red. M class drug codes are presented in Table~\ref{tab:codes_m_drugs}. }
    \label{tbl:table_of_figures}
\end{table*}

\section{Discussion}

Our study showed that tree-based ML models trained on linked administrative datasets, in tandem with XAI techniques, have the potential to act as an ‘early warning system’ for per-patient ACS related adverse outcomes and for harmful drugs. To the best of our knowledge, this is the first study to demonstrate this using XAI.

The results of our feature importance analysis show that sex and age are ranked highly by MDI, MDA, and SHAP --- which is expected, as these are the most important confounding variables in clinical and epidemiological studies. Certain C class drugs (e.g. digoxin, glyceryl trinitrate, isosorbide mononitrate, and furosemide) and comorbidities (see Figure~\ref{tbl:table_of_feat_imps} and Appendix~\ref{a:lookuptables}) are consistently reported as being highly important across all feature importance measures, which suggests some level of model independent feature importance. Importantly, the M class drugs celecoxib, rofecoxib, and alendronic acid repeatedly appear as important features across MDI, MDA, and SHAP analyses. 

The LIME feature importance scores are inconsistent with the other measures for certain features (see Figure~\ref{tbl:table_of_feat_imps}). A potential explanation is the variance across multiple consecutive runs in LIME \cite{alvarez2018robustness} --- features which were ranked highly in one run could be ranked lower in another \cite{man2020finance}. Despite this, LIME has been shown to be at least as stable as SHAP for highly ranked features \cite{dieber2020model}. Our results agree with this: the highly ranked comorbidity features are consistently ranked highly across all feature importance measures.

Random features were included because by definition they cannot have any predictive power --- nevertheless, the MDI analysis ranked these features as important. MDI measures have a bias whereby variables with a large number of categories or potential values (high cardinality) are ranked as more important \cite{altmann2010permutation}. It is thereby expected that an impurity based approach will inflate the importance of random features. Moreover, MDI feature importance scores are calculated over the decision trees generated by the training dataset, so the importance of non-predictive features may be inflated. Under MDA, these high feature importance scores are reduced as the permutation importance scores are computed over a held out test set (Figure~\ref{tbl:table_of_feat_imps}). 

Each of our feature importance analyses demonstrate that celecoxib and rofecoxib are important when predicting ACS based adverse outcomes, or more specifically, that they actually \edit{provide a positive probability contribution} to ACS based adverse outcome predictions (see Figure~\ref{tbl:table_of_feat_imps}). For MDI and MDA, celecoxib and rofecoxib are ranked as the most important M class drugs when making ACS based adverse outcomes predictions. LIME and SHAP analysis shows that both drugs contribute to ACS related adverse outcomes predictions, and SHAP analysis further ranks them as the second and third \textit{most} contributing M class drugs. \edit{However, LIME analysis heavily mutes the importance of these features compared to SHAP, greatly reducing the magnitude of their on-average contributions}. 

\edit{Our stratified analysis (see Figure~\ref{tbl:table_of_figures} and Table~\ref{tab:permodelmclass}) show more detailed results: having \textit{any} celecoxib dispensings in an individual's drug history increases the likelihood of an ACS related adverse outcome prediction (M01AH01 / celecoxib's contribution is positive in Figure~\ref{tbl:table_of_figures} rows a and b), and \textit{not having any} celecoxib dispensings decreases or does not impact the likelihood of an ACS related adverse outcome prediction (M01AH01 / celecoxib's contribution is negative or near-zero in Figure~\ref{tbl:table_of_figures} rows c and d). In other words, the presence of celecoxib dispensing events increases the likelihood of a positive ACS prediction independent of rofecoxib dispensing events.}

\edit{This  is the equivalent of the counterfactual in the epidemiological theory of causality. The counterfactual conditionals are of the form ``if A had not occurred, then C would not have occurred''. This is exactly what is observed in Figure~\ref{tbl:table_of_figures}. These results provide more robust epidemiological evidence of the associations of celecoxib and rofecoxib with the outcomes.}

We acknowledge the utility of existing statistical methods in pharmacovigilance. For example, Sequence Symmetry Analysis (SSA) has been determined to be a promising solution when compared to other quantitative methods such as Reporting Odds Ratio (ROR), Proportional Reporting Ratio (PRR), and Bayesian techniques \cite{curtis2008bayesian,wahab2013adrcoxib}. SSA considers the distribution of diseases and drugs, before and after initiation of treatment for an adverse event. Asymmetry in these distributions indicate that adverse events may be due to drug supply \cite{tsiropoulos2009antiepileptic}. SSA's time to signal detection for rofecoxib-induced myocardial infarction was within 1-3 years of market entry, whereas it took 5-7 years after market entry for trial results to lead to warnings and withdrawals of the drugs \cite{wahab2013adrcoxib}. Despite this, there are limitations: in these investigations there are no adverse event signals which can be linked from a joint database; the signals are deduced from dispensing data only --- e.g. furosemide initiation being used as an indicator of heart failure \cite{wahab2016ssahf}. Using joint databases is more reliable as administrative data entries can validate if a comorbidity did actually occur. Moreover, XAI is better suited for domains which are already using ML-based approaches.

The administrative data used in this study has limitations. By definition, administrative data only includes information which is pertinent to administration. Clinical features which may impact the likelihood of a patient having an ACS related adverse outcome may not be present, and rich clinical information is reduced to ICD-10-AM codes. Also, there were assumptions which we had to apply --- for example, we assumed that patients who are dispensed a given drug actually consume it. These shortcomings may have impacted the predictive performance of our models (Table~\ref{tab:modelscore}). Moreover, the RF, ET, and XGB in this study converged to similar performances, suggesting that there may be a limit to the predictive performance which can be attained using these datasets. \edit{Indeed, other ML based predictors trained on administrative data have reached similar performance limits \cite{han2019spine, brignone2018applying}.} 

\edit{Further potential research to progress the application of ML techniques in pharmacovigilance include the testing of more diverse ML models (and studies into their inductive biases)}, benchmarking these methods against other datasets that include more clinical data, investigating additional XAI algorithms \cite{ribeiro2018anchors}, using the feature importance scores to iteratively reduce the full set of features, and handling multicollinear features whilst maintaining interpretability. \edit{Counter-factual examples may represent an interesting XAI approach to providing transparency in the pharmacovigilance domain, but calculating such explanations may be intractable for high-dimensionality datasets \cite{ramon2020shapclimec}.}

The strength of our study was the use of multiple linked administrative datasets that cover the whole-of-population. Another strength was the \textit{validation} of feature importances scoring methods against features that have somewhat 'known' importances (e.g., it is known that rofecoxib increases the risk of ACS, and that random features are unimportant). \edit{Ultimately, we believe that this work highlights the utility of XAI in further analysing trained AI models in challenging domains, and in cases where performance may be constrained due to dataset limitations.} 

\subsection{Conclusion}

Tree-based ML models trained on linked administrative datasets, in tandem with XAI techniques, have the potential to act as an `early warning system' for per-patient ACS related adverse outcomes and for harmful drugs in a pharmacovigilance monitoring system. MDA and SHAP methods exceed MDI and LIME methods in identifying known harmful drugs, key confounding variables, and random features. With the appropriate infrastructure and additional clinical data, these algorithms could provide an autonomous method of monitoring adverse outcomes from medications at the population level. This represents a valuable addition to the existing statistical techniques which are currently used and would be the next step in progressing towards a real-time pharmacovigilance monitoring system.

\subsection{Author contributions}

FM.S., G.D. conceived the study and provided clinical interpretation. IR.W. prepared the data based on a data preparation approach originally implemented by J.L.. IR.W conducted the experiments. M.B. contributed to the design of machine learning and the use of explainable artificial intelligence. J.L. and L.W. contributed to fortnightly discussions which were led by FM.S., which shaped this work. IR.W, FM.S, J.L, and L.W have access to the underlying data and have verified it. FM.S. contributed to the conception, study design, data acquisition, funding, analysis, and supervision of IR.W.. All co-authors critically revised the work and gave final approval and agreed to be accountable for all aspects of the work ensuring integrity and accuracy.

\subsection{Ethics statement}

This study complies with the Declaration of Helsinki. Human Research Ethics Committee approvals were received from the Western Australian Department of Health (\#2011/62); the Australian Department of Health (XJ-16); and the University of Western Australia (RA/4/1/1130).

\subsection{Data sharing statement}

We will consider requests for data sharing on an individual basis, with the aim to share data whenever possible for appropriate research purposes. However, this research project uses data obtained from a third-party source under strict privacy and confidentiality agreements from the Western Australian Department of Health (State) and Australian Department of Health (Federal) databases, which are governed by their ethics committees and data custodians. The data were provided after approval was granted from their standard application processes for access to the linked datasets. Therefore, any requests to share these data with other researchers will be subject to formal approval from the third-party ethics committees and data custodian(s). Researchers interested in these data should contact the Data Services Team at the Data Linkage Branch of the Western Australian Department of Health (www.datalinkage-wa.org.au/contact-us).

\subsection{Competing interests}

Girish Dwivedi reports personal fees from Pfizer, Amgen, Astra Zeneca and Artrya Pty Ltd, all of which are outside the submitted work. No other competing interests were declared.

\begin{acks}
    We thank the following institutions and groups for providing the data used in this study: staff at the WA Data Linkage Branch and data custodians of the WA Department of Health Inpatient Data Collections and Registrar General for access to and provision of the State linked data; the Australian Department of Health for the cross-jurisdictional linked PBS data; the Victorian Department of Justice and Regulation for the cause of death data held in the National Coronial Information System. We also thank the University of WA for funding this work through the 2019 FHMS research grant.
\end{acks}

% -------------------------------------------------------------
% FIN. --------------------------------------------------------
% -------------------------------------------------------------

%\bibliographystyle{acm}
\bibliographystyle{naturemag}
\bibliography{main}

\appendix

\newpage
\section{Feature Lookup tables and code descriptions}
\label{a:lookuptables}

\begin{table}[H]
    \caption{ Descriptions of the features `sex', `age', `rand float', and `rand int'. 
    }
    \vspace{-2mm}
    \centering
    \begin{tabular}{p{1.2cm} p{6.8cm}}
        \hline
        Feature  & Description \\
        \hline
        sex &  The individual's sex. The values '1', and '2' represent male and female. It was not possible to consider other values in this study, due to their low sample size.  \\
        age &  The individual's age in years. \\
        rand float & A random floating point number in the range 0 --- 1. \\
        rand int & A random integer in the range 0 --- 157 inclusive (the number of features considered in the study). \\
        \hline
    \end{tabular}
    \label{tab:codes_general}
    \vspace{-6mm}
\end{table}

\begin{table}[H]
    \centering
    \caption{ Drug names of the ATC codes for the C class drugs C01AA05 --- C09AA10. In this study, the features corresponding to these codes represent the number of dispensings that an individual had for these drugs in the drug history period. }
    %\vspace{-2mm}
    \begin{tabular}{p{1.4cm} p{6.6cm}}
        \hline
        ATC code  & Drug name \\
        \hline
        C01AA05 & digoxin \\
        C01BC04 & flecainide \\
        C01BD & Antiarrhythmics, class III \\
        C01BD01 & amiodarone \\
        C01DA02 & glyceryl trinitrate \\
        C01DA14 & isosorbide mononitrate \\
        C01DX16 & nicorandil \\
        C02AB01 & \textit{l}-methyldopa \\
        C02AC01 & clonidine \\
        C02CA01 & prazosin \\
        C03AA01 & bendroflumethiazide \\
        C03AA03 & hydrochlorothiazide \\
        C03BA11 & indapamide \\
        C03CA01 & furosemide \\
        C03DA01 & spironolactone \\
        C03DB01 & amiloride \\
        C03EA01 & hydrochlorothiazide \& potassium-sparing agents \\
        C07AA03 & pindolol \\
        C07AA05 & propranolol \\
        C07AB02 & metoprolol \\
        C07AB03 & atenolol \\
        C07AB07 & bisoprolol \\
        C07AG02 & carvedilol \\
        C08CA01 & amlodipine \\
        C08CA02 & felodipine \\
        C08CA05 & nifedipine \\
        C08CA13 & lercanidipine \\
        C08DA01 & verapamil \\
        C08DB01 & diltiazem \\
        C09AA01 & captopril \\
        C09AA02 & enalapril \\
        C09AA03 & lisinopril \\
        C09AA04 & perindopril \\
        C09AA05 & ramipril \\
        C09AA06 & quinapril \\
        C09AA09 & fosinopril \\
        C09AA10 & trandolapril \\
        \hline
    \end{tabular}
    \label{tab:codes_c_drugs_a}
\end{table}

\begin{table}[H]
    \centering
    \caption{ Drug names of the ATC codes for the C class drugs C09BA02 --- C10BX03. In this study, the features corresponding to these codes represent the number of dispensings that an individual had for these drugs in the drug history period. }
    \begin{tabular}{p{1.4cm} p{6.6cm}}
        \hline
        ATC code  & Drug name \\
        \hline
        C09BA02 & enalapril and diuretics \\
        C09BA04 & perindopril and diuretics \\
        C09BA06 & quinapril and diuretics \\
        C09BA09 & fosinopril and diuretics \\
        C09CA02 & eprosartan \\
        C09CA04 & irbesartan \\
        C09CA06 & candesartan \\
        C09CA07 & telmisartan \\
        C09DA02 & eprosartan and diuretics \\
        C09DA04 & irbesartan and diuretics \\
        C09DA06 & candesartan and diuretics \\
        C09DA07 & telmisartan and diuretics \\
        C10AA01 & simvastatin \\
        C10AA03 & pravastatin \\
        C10AA04 & fluvastatin \\
        C10AA05 & atorvastatin \\
        C10AA07 & rosuvastatin \\
        C10AB04 & gemfibrozil \\
        C10AB05 & fenofibrate \\
        C10AX09 & ezetimibe \\
        C10BA02 & simvastatin and ezetimibe \\
        C10BX03 & atorvastatin and amlodipine \\
        \hline
    \end{tabular}
    \label{tab:codes_c_drugs_b}
\end{table}

\begin{table}[H]
    \centering
    \caption{ Drug names of the ATC codes for all the M class drugs considered in this study. In this study, the features corresponding to these codes represent the number of dispensings that an individual had for these drugs in the drug history period. }
    \begin{tabular}{p{1.4cm} p{6.6cm}}
        \hline
        ATC code  & Drug name \\
        \hline
        M01AB01 & indometacin \\
        M01AB05 & diclofenac \\
        M01AC01 & piroxicam \\
        M01AC06 & meloxicam \\
        M01AE01 & ibuprofen \\
        M01AE02 & naproxen \\
        M01AE03 & ketoprofen \\
        M01AH01 & celecoxib \\
        M01AH02 & rofecoxib \\
        M01AH06 & lumiracoxib \\
        M01CA & Quinolines, antirheumatic drugs \\
        M03BX01 & baclofen \\
        M04AA01 & allopurinol \\
        M04AC01 & colchicine \\
        M05BA04 & alendronic acid \\
        M05BA07 & risedronic acid \\
        M05BB02 & risedronic acid and calcium, sequential \\
        M05BB03 & alendronic acid and colecalciferol \\
        M05BX & Other drugs affecting bone structure and mineralization in ATC \\
        \hline
    \end{tabular}
    \label{tab:codes_m_drugs}
\end{table}

\begin{table}[H]
    \centering
    \caption{ Descriptions of the ICD-10-AM codes I20.0 --- I50.9. In this study, the features corresponding to these codes represent the number of hospitalisations that an individual had due to these conditions in the comorbidity history period. }
    \begin{tabular}{p{1.0cm} p{7.0cm}}
        \hline
        ICD-10  & Description \\
        \hline
        I20.0 &  Unstable angina \\
        I20.1 &    Angina pectoris with documented spasm \\
        I20.8 &    Other forms of angina pectoris \\
        I20.9 &    Angina pectoris, unspecified \\
        I21.0 &  ST elevation myocardial infarction of anterior wall \\
        I21.1 &  ST elevation myocardial infarction of inferior wall \\
        I21.2 &  ST elevation myocardial infarction of other sites \\
        I21.3 &    ST elevation myocardial infarction of unspecified site \\
        I21.4 &    Non-ST elevation myocardial infarction \\
        I21.9 &    Acute myocardial infarction, unspecified \\
        I22.0 &  Subsequent ST elevation and non-ST elevation myocardial infarction \\
        I22.1 &    Subsequent ST elevation myocardial infarction of inferior wall \\
        I22.8 &    Subsequent ST elevation myocardial infarction of other sites \\
        I22.9 &    Subsequent ST elevation myocardial infarction of unspecified site \\
        I23.0 &  Certain current complications following ST elevation and non-ST elevation myocardial infarction (within the 28 day period) \\
        I23.1 &    Atrial septal defect as current complication following acute myocardial infarction \\
        I23.2 &    Ventricular septal defect as current complication following acute myocardial infarction \\
        I23.3 &    Rupture of cardiac wall without hemopericardium as current complication following acute myocardial infarction \\
        I23.5 &    Rupture of papillary muscle as current complication following acute myocardial infarction \\
        I23.6 &    Thrombosis of atrium, auricular appendage, and ventricle as current complications following acute myocardial infarction \\
        I23.8 &    Other current complications following acute myocardial infarction \\
        I24.0 &  Other acute ischemic heart diseases \\
        I24.1 &    Dressler's syndrome \\
        I24.8 &    Other forms of acute ischemic heart disease \\
        I24.9 &    Acute ischemic heart disease, unspecified \\
        I25.0 &   Chronic ischemic heart disease \\
        I25.1 &  Atherosclerotic heart disease of native coronary artery \\
        I25.2 &    Old myocardial infarction \\
        I25.3 &    Aneurysm of heart \\
        I25.4 &  Coronary artery aneurysm and dissection \\
        I25.5 &    Ischemic cardiomyopathy \\
        I25.6 &    Silent myocardial ischemia \\
        I25.8 &  Other forms of chronic ischemic heart disease \\
        I25.9 &    Chronic ischemic heart disease, unspecified \\
        I50.0 &  Heart failure \\
        I50.1 &    Left ventricular failure, unspecified \\
        I50.9 &    Heart failure, unspecified \\
        \hline
    \end{tabular}
    \label{tab:codes_icd_a}
\end{table}

\begin{table}[H]
    \centering
    \caption{ Descriptions of the ICD-10-AM codes I60.0 --- I64, K92.0 -- K92.2, and R58. In this study, the features corresponding to these codes represent the number of hospitalisations that an individual had due to these conditions in the comorbidity history period.}
    \begin{tabular}{p{1.0cm} p{7.0cm}}
        \hline
        ICD-10  & Description \\
        \hline
        I60.0 &   Nontraumatic subarachnoid hemorrhage from unspecified carotid siphon and bifurcation \\
        I60.1 &   Nontraumatic subarachnoid hemorrhage from unspecified middle cerebral artery \\
        I60.2 &    Nontraumatic subarachnoid hemorrhage from anterior communicating artery \\
        I60.3 &   Nontraumatic subarachnoid hemorrhage from unspecified posterior communicating artery \\
        I60.4 &    Nontraumatic subarachnoid hemorrhage from basilar artery \\
        I60.5 &   Nontraumatic subarachnoid hemorrhage from unspecified vertebral artery \\
        I60.6 &    Nontraumatic subarachnoid hemorrhage from other intracranial arteries \\
        I60.7 &    Nontraumatic subarachnoid hemorrhage from unspecified intracranial artery \\
        I60.8 &    Other nontraumatic subarachnoid hemorrhage \\
        I60.9 &    Nontraumatic subarachnoid hemorrhage, unspecified \\
        I61.0 &  Nontraumatic intracerebral hemorrhage in hemisphere, subcortical \\
        I61.1 &    Nontraumatic intracerebral hemorrhage in hemisphere, cortical \\
        I61.2 &    Nontraumatic intracerebral hemorrhage in hemisphere, unspecified \\
        I61.3 &    Nontraumatic intracerebral hemorrhage in brain stem \\
        I61.4 &    Nontraumatic intracerebral hemorrhage in cerebellum \\
        I61.5 &    Nontraumatic intracerebral hemorrhage, intraventricular \\
        I61.6 &    Nontraumatic intracerebral hemorrhage, multiple localized \\
        I61.8 &    Other nontraumatic intracerebral hemorrhage \\
        I61.9 &    Nontraumatic intracerebral hemorrhage, unspecified \\
        I62.0 &   Nontraumatic subdural hemorrhage, unspecified \\
        I62.1 &    Nontraumatic extradural hemorrhage \\
        I62.9 &    Nontraumatic intracranial hemorrhage, unspecified \\
        I63.0 &   Cerebral infarction due to thrombosis of unspecified precerebral artery \\
        I63.1 &   Cerebral infarction due to embolism of unspecified precerebral artery \\
        I63.2 &   Cerebral infarction due to unspecified occlusion or stenosis of unspecified precerebral arteries \\
        I63.3 &   Cerebral infarction due to thrombosis of unspecified cerebral artery \\
        I63.4 &   Cerebral infarction due to embolism of unspecified cerebral artery \\
        I63.5 &   Cerebral infarction due to unspecified occlusion or stenosis of unspecified cerebral artery \\
        I63.6 &    Cerebral infarction due to cerebral venous thrombosis, nonpyogenic \\
        I63.8 &  Other cerebral infarction \\
        I63.9 &    Cerebral infarction, unspecified \\
        I64 &  Stroke, not specified as haemorrhage or infarction \\
        \hline
        K92.0 &  Hematemesis \\
        K92.1 &    Melena \\
        K92.2 &    Gastrointestinal hemorrhage, unspecified \\
        R58 &     Hemorrhage, not elsewhere classified \\
        \hline
    \end{tabular}
    \label{tab:codes_icd_b}
\end{table}

\newpage
\section{Hyperparameter Searches}
\label{a:hyper}

\begin{table}[H]
    \centering
    \caption{ The range of values that the hyperparameters could take during our RF and ET hyperparameter optimisation. }
    \begin{tabular}{p{2.6cm} p{5.4cm}}
        \hline
        Hyperparameter  & Search distribution \\
        \hline
        n\textunderscore estimators & 8 --- 32 \\
        max\textunderscore features & `auto', `sqrt' \\
        max\textunderscore depths & 8 --- 64, no limit \\
        min\textunderscore samples\textunderscore split &   2 --- 12 \\
        min\textunderscore samples\textunderscore leaf &   2 --- 8 \\
        bootstrap &   True, False \\
        \hline
    \end{tabular}
    \label{tab:hyper_rf_et}
\end{table}

\begin{table}[H]
    \centering
    \caption{ The range of values that the hyperparameters could take during our XGB hyperparameter optimisation. }
    \begin{tabular}{p{2.6cm} p{5.4cm}}
        \hline
        Hyperparameter  & Search distribution \\
        \hline
        booster &   `gbtree', `gblinear' \\
        max\textunderscore depth &   2 --- 32, no limit \\
        sampling\textunderscore method &  `uniform', `gradient\textunderscore based' \\
        alpha &  0, 0.1, 0.5 \\
        lambda &   0.8, 1, 1.2 \\
        grow\textunderscore policy &   `depthwise', `lossguide'  \\
        \hline
    \end{tabular}
    \label{tab:hyper_xgb}
\end{table}

\begin{comment}
clf_rfc.get_params()
{'bootstrap': False,
 'ccp_alpha': 0.0,
 'class_weight': None,
 'criterion': 'gini',
 'max_depth': 44,
 'max_features': 'sqrt',
 'max_leaf_nodes': None,
 'max_samples': None,
 'min_impurity_decrease': 0.0,
 'min_impurity_split': None,
 'min_samples_leaf': 2,
 'min_samples_split': 7,
 'min_weight_fraction_leaf': 0.0,
 'n_estimators': 180,
 'n_jobs': None,
 'oob_score': False,
 'random_state': None,
 'verbose': 0,
 'warm_start': False}
 
clf_etc.get_params()
{'bootstrap': False,
 'ccp_alpha': 0.0,
 'class_weight': None,
 'criterion': 'gini',
 'max_depth': 69,
 'max_features': 'sqrt',
 'max_leaf_nodes': None,
 'max_samples': None,
 'min_impurity_decrease': 0.0,
 'min_impurity_split': None,
 'min_samples_leaf': 2,
 'min_samples_split': 7,
 'min_weight_fraction_leaf': 0.0,
 'n_estimators': 350,
 'n_jobs': None,
 'oob_score': False,
 'random_state': None,
 'verbose': 0,
 'warm_start': False}
 
clf_xgb.get_params()
{'ccp_alpha': 0.0,
 'criterion': 'friedman_mse',
 'init': None,
 'learning_rate': 0.1,
 'loss': 'deviance',
 'max_depth': 69,
 'max_features': 'sqrt',
 'max_leaf_nodes': None,
 'min_impurity_decrease': 0.0,
 'min_impurity_split': None,
 'min_samples_leaf': 2,
 'min_samples_split': 7,
 'min_weight_fraction_leaf': 0.0,
 'n_estimators': 350,
 'n_iter_no_change': None,
 'presort': 'deprecated',
 'random_state': None,
 'subsample': 1.0,
 'tol': 0.0001,
 'validation_fraction': 0.1,
 'verbose': 0,
 'warm_start': False}
\end{comment}

\end{document}